\documentclass[runningheads]{llncs}
\usepackage[T1]{fontenc}
\usepackage{graphicx,verbatim}
\usepackage{cite}
\usepackage{algorithm}
\usepackage{algpseudocode}%

\usepackage{amsmath}
\usepackage{amsmath} 
\usepackage{amssymb}
\usepackage{url}
\usepackage{breakurl}
\usepackage{hyperref}
\DeclareMathOperator*{\median}{median}

\usepackage[misc]{ifsym}

\usepackage{bm}
\usepackage{booktabs}
\usepackage{makecell}  
\usepackage{array,color}
\usepackage{array}
\usepackage{multirow}

\usepackage[misc]{ifsym}
\usepackage{multirow}  
\usepackage{booktabs}
\usepackage{threeparttable}
\usepackage{bbding}
\usepackage{ulem}
\usepackage{pifont}

\usepackage{tabularx}

\begin{document}
\bibliographystyle{splncs04}
\title{Shape-aware Sampling Matters in the Modeling of Multi-Class Tubular Structures}
\titlerunning{Shape-aware Sampling}
\author{
Minghui Zhang\textsuperscript{*}\inst{1,2} \and 
Yaoyu Liu\textsuperscript{*}\inst{1,2}   \and 
Xin You \inst{1,2} \and 
Hanxiao Zhang\inst{1}  \and 
Yun Gu\inst{1,2}\textsuperscript{(\Letter)}
}

\renewcommand{\thefootnote}{\fnsymbol{footnote}}
\footnotetext[1]{Equal contribution.}

\authorrunning{M. Zhang et al.}
\institute{Institute of Medical Robotics, Shanghai Jiao Tong University, Shanghai, China
\email{
   \{minghuizhang, geron762\}@sjtu.edu.cn}\\
\and
Department of Automation, Shanghai Jiao Tong University, Shanghai, China
}
\maketitle              

\begin{abstract}
Accurate multi-class tubular modeling is critical for precise lesion localization and optimal treatment planning. 
Deep learning methods enable automated shape modeling by prioritizing volumetric overlap accuracy.
However, the inherent complexity of fine-grained semantic tubular shapes is not fully emphasized by overlap accuracy, resulting in reduced topological preservation.
To address this, we propose the Shape-aware Sampling (SAS), which optimizes patchsize allocation for online sampling and extracts a topology-preserved skeletal representation for the objective function.
Fractal Dimension-based Patchsize (FDPS) is first introduced to quantify semantic tubular shape complexity through axis-specific fractal dimension analysis. 
Axes with higher fractal complexity are then sampled with smaller patchsizes to capture fine-grained features and resolve structural intricacies.
In addition, Minimum Path-Cost Skeletonization (MPC-Skel) is employed to sample topologically consistent skeletal representations of semantic tubular shapes for skeleton-weighted objective functions. 
MPC-Skel reduces artifacts from conventional skeletonization methods and directs the focus to critical topological regions, enhancing tubular topology preservation.
SAS is computationally efficient and easily integrable into optimization pipelines. Evaluation on two semantic tubular datasets showed consistent improvements in both volumetric overlap and topological integrity metrics.
\end{abstract}

\section{Introduction}
\begin{figure*}[t]
\centering
\includegraphics[width=0.95\linewidth]{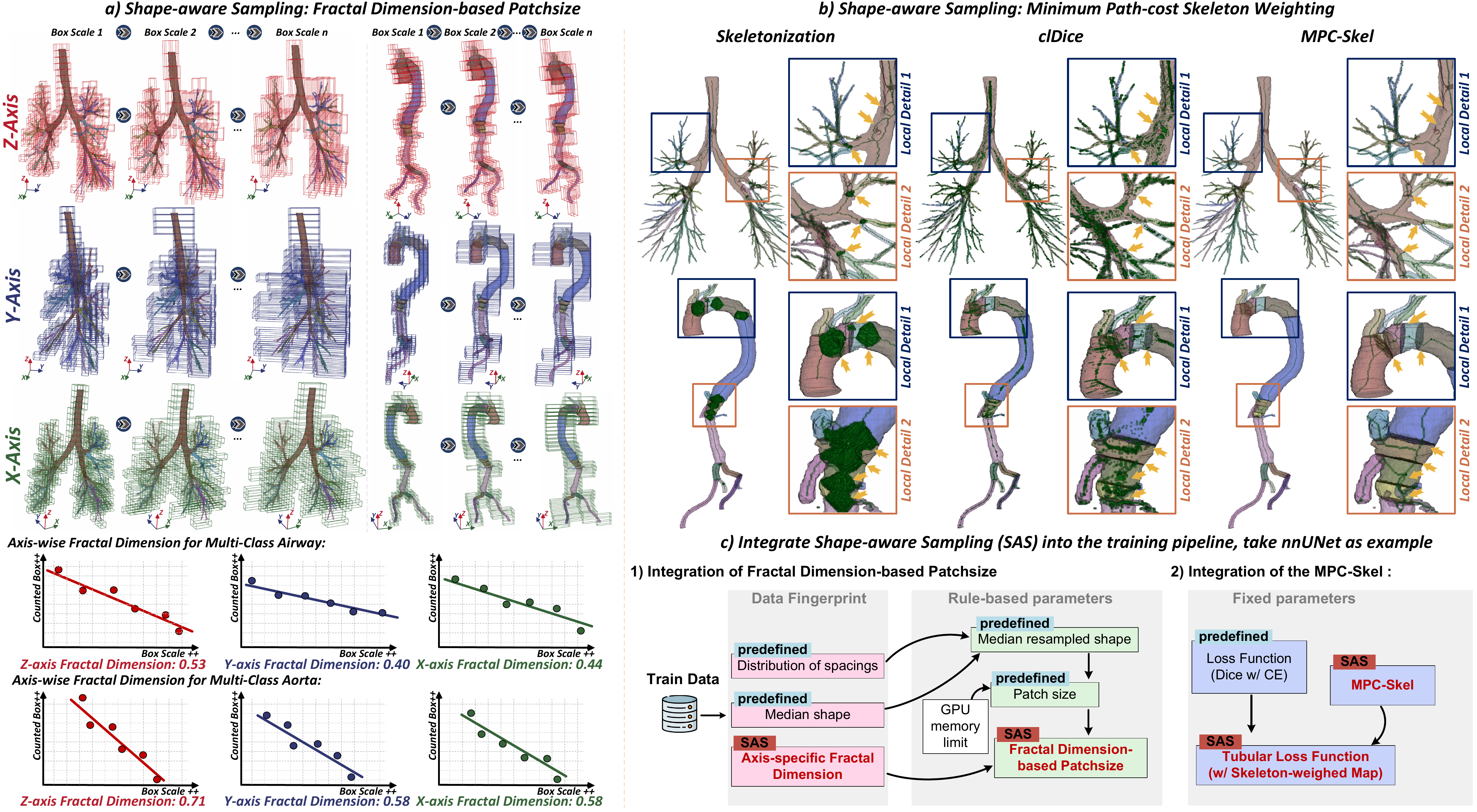}
\caption{The framework of the Shape-aware Sampling (SAS).
a) FDPS quantifies the axis-specific fractal dimension and subsequently reassigns the patchsize based on axis-wise shape complexity.
b) MPC-Skel samples the topology-preserved skeletal representation for precise skeleton-based weighting.
c) SAS is computationally efficient and can be seamlessly integrated into training pipelines.}
\label{fig:framework}
\end{figure*}

Fine-grained modeling of semantic tubular structures enables accurate identification of pathological lesions within their specific anatomical context, thereby facilitating the development of optimized diagnostic and therapeutic planning 
strategies. For example, anatomical structures like the aorta and bronchi can be segmented into clinically relevant fine-grained branches. Modeling the multi-class aorta \cite{imran2025multi, imran2024cis} allows for the measurement of the volumes and 
radii of various aortic branches, facilitating the assessment of dissection extent and informing optimal stent placement\cite{carrel2023acute,rolf2024mechanisms}. Anatomical labeling of multi-class bronchi \cite{yu2022tnn} aids in identifying 
the local branches affected by pathological lesions, providing a detailed preoperative planning map for endoscopic navigation surgeries \cite{zhang2023soft}.
Recent advances in deep learning have enabled automated volumetric modeling of medical structures \cite{cciccek20163d, isensee2021nnu}. 
While existing approaches predominantly emphasize the optimization of volumetric overlap metrics, demonstrating efficacy in modeling of human organs, 
they exhibit limitations when applied to tubular structures. This limitation arises because volumetric overlap measures cannot maintain the crucial topology integrity required for tubular structures.
Moreover, the inherent complexity of multi-class tubular structures presents additional challenges due to their heterogeneous spatial configurations across different anatomical axes. 

To alleviate this challenge, we propose the Shape-aware Sampling (SAS) to enhance the modeling of semantic tubular structures. 
SAS customizes patchsize sampling based on heterogeneous spatial shape complexity and samples topology-preserved skeletal representation to emphasize topology integrity.
As for the patchsize sampling, it is typically determined by the median resampled shape while considering GPU memory constraints \cite{isensee2021nnu}. However, this approach overlooks the varying complexity of the shape across different axes.
Recent studies \cite{isensee2024nnu, isensee2024scaling} suggest using larger patchsizes to better capture spatial relationships in multi-class objects. However, this significantly increases memory consumption and training time.
To overcome this, SAS introduces the Fractal Dimension-based Patchsize (FDPS) strategy to efficiently optimize patchsize allocation during training. 
As illustrated in Fig. \ref{fig:framework}.(a), FDPS assesses heterogeneous shape complexity via axis-specific fractal dimension analysis \cite{konatar2020box}. 
A higher fractal dimension reflects increased structural intricacy and finer details along the axis \cite{falconer2013fractal}.
After initial patchsize configuration, FDPS dynamically reassigns the patchsize inversely proportional to the fractal dimension of each axis. 
The axis with greater fractal complexity samples smaller patchsize, improving the capability of the network to resolve local details and capture fine-grained features.

In addition, recent studies \cite{zhang2023towards, SkeletonRecall, shi2024centerline} have introduced skeleton-weighted loss functions to enhance topology-aware segmentation of thin tubular structures. 
However, the commonly used skeletonization techniques \cite{lee1994building,shit2021cldice} often generate false positive noises, manifesting as clump-like or spurious structures \cite{attali2009stability, cornea2024curve}, as shown in Fig.\ref{fig:framework}.(b). 
These artifacts distort the calculation of the skeleton-weighted map, resulting in an inaccurate focus for the topology of semantic tubular shapes.
To mitigate this, SAS implements a Minimum Path-Cost Skeletonization (MPC-Skel) that generates topology-preserved skeletal representation for precise skeleton-weighting. 
Specifically, MPC-Skel iteratively samples the shortest paths within the volume, guided by a cost function derived from distance transformation. After each trace, an adaptive sphere expands the voxel set near the path. 
The minimum path-cost trace and adaptive expansion helps mitigate false positive noises, as illustrated in Fig. \ref{fig:framework}.(b). 
In contrast to \cite{lee1994building, shit2021cldice}, MPC-Skel produces a more accurate skeleton-weighted map, significantly improving the preservation of tubular topology. 
Both FDPS and MPC-Skel are computationally efficient and easily integrable into training pipelines, as shown in Fig. \ref{fig:framework}.(c). A comprehensive evaluation on two multi-class tubular datasets, 
using three different backbones, demonstrates that SAS consistently enhances both volumetric overlap and topological integrity.

\section{Method}
In this section, we detail the proposed Shape-aware Sampling (SAS). The overall framework is illustrated in Fig. \ref{fig:framework}. SAS comprises the Fractal Dimension-based Patchsize (FDPS) 
and Minimum Path-Cost Skeletonization (MPC-Skel). The integration of SAS into the training pipeline is demonstrated in Algorithm 1. 

\subsection{Shape-aware Sampling: Fractal Dimension based Patchsize}
While increasing the patchsize improves segmentation performance, it concurrently results in a significant increase in training time \cite{isensee2024nnu, isensee2024scaling, imran2025multi}. 
Given that the complexity of the semantic tubular structures varies among different axes, we first introduce the fractal dimension to measure the axis-wise geometric complexity \cite{falconer2013fractal, pentland1984fractal}.
Based on the axis-specific shape complexity, SAS introduces the Fractal Dimension-based Patchsize (FDPS) to optimize patchsize allocation for online sampling. 
FDPS customizes the patchsize based on the axis-specific complexity of the fractal dimension, without increasing the training duration.
The box-counting method \cite{liebovitch1989fast, li2009improved, konatar2020box} is employed to approximate the fractal dimension of volumetric shapes. 
For a given volumetric shape, we count the number of the boxes $N(r)$ that required to cover the object at the certain size of $r$. 
This relationship can be descibed as $N(r) \backsim r^{-FD}$, where $\mathnormal{FD}$ denotes the fractal dimension. 
Furthermore, we compute the fractal dimension along each axis, and perform box-counting at several different scales $r$. The axis-specific fractal dimension is then obtained as:
\begin{align}
\mathnormal{FD}_{i} = \lim_{r_{i} \to 0} \frac{\log N(r_{i})}{-\log r_{i}}, i = \{x, y, z\}. \label{eq:Axis-wise_FD}
\end{align}
Here, $\mathnormal{FD}_{x}$, $\mathnormal{FD}_{y}$, $\mathnormal{FD}_{z}$ represent the fractal dimension along the $x-$, $y-$, and $z-$ axes, respectively. 
The complexity of the fractal dimension can be ranked by: $\mathnormal{FD}_{\text{max}}$, $\mathnormal{FD}_{\text{mid}}$, $\mathnormal{FD}_{\text{min}}$ = $\max$, $\median$, $\min$ $(\mathnormal{FD}_{x},\mathnormal{FD}_{y},\mathnormal{FD}_{z})$. 
Similarly, $\mathnormal{PS}_{x}$, $\mathnormal{PS}_{y}$, and $\mathnormal{PS}_{z}$ denote the initial patchsize configuration, and 
$\mathnormal{PS}_{\text{max}}$, $\mathnormal{PS}_{\text{mid}}$, $\mathnormal{PS}_{\text{min}}$ =  $\max$, $\median$, $\min$ $(\mathnormal{PS}_{x},\mathnormal{PS}_{y},\mathnormal{PS}_{z})$. 
We hypothesize that the patchsize for each axis during training should consider not only the size of the shape but also the heterogeneous complexity. 
A higher fractal dimension reflects increased structural intricacy and finer details along the axis. Therefore, FDPS reassigns the patchsize inversely proportional to the fractal dimension:
\begin{align}
\mathnormal{FDPS}_i = \begin{cases} 
\mathnormal{PS}_{\text{min}}, & \text{if } \mathnormal{FD}_{i} = \mathnormal{FD}_{\text{max}} \\
\mathnormal{PS}_{\text{mid}}, & \text{if } \mathnormal{FD}_{i} = \mathnormal{FD}_{\text{mid}} \\
\mathnormal{PS}_{\text{max}}, & \text{if } \mathnormal{FD}_{i} = \mathnormal{FD}_{\text{min}}
\end{cases}, i = \{x, y, z\}. \label{eq:FDPS}
\end{align}
In this approach, axes with greater fractal complexity are sampled with smaller patchsizes, which enhances the ability of the network to resolve local structural details and capture fine-grained features.

\begin{algorithm}[t]
    \caption{Shape-aware Sampling embedded in the training pipeline.}
    \begin{algorithmic}[1]
        \Require target label ($\mathnormal{Y}$), initial patchsize ($\mathnormal{PS}_{x}$,$\mathnormal{PS}_{y}$,$\mathnormal{PS}_{z}$), network prediction ($\hat{\mathnormal{Y}}$)
        \For {$i \in \{x, y, z\}$}
            \State $List_{N(r)}$ $\gets [\,]$
            \For{$r \gets 2$ to $ \text{Size}_{i}(\mathnormal{Y})/ 2$}
            \State Split $\mathnormal{Y}$ to separate boxes with size of $r$;
            \State $N(r) \gets$ Number of counted boxes that contains foreground;
             \State $List_{N(r)}$ append ($log(N(r)), -log(r)$);
            \EndFor
            \State $\mathnormal{FD}_{i} \gets$ \textbf{LinearRegression}($List_{N(r)}$)
        \EndFor
        \State $FDPS_{\{x,y,z\}}\gets $ \textbf{Rank\&Reassign}($\mathnormal{PS}_{\{x,y,z\}}$, $\mathnormal{FD}_{\{x,y,z\}}$) \hfill \textit{\% using Eq. 2}
        \State $\mathnormal{Y}_{MPC-Skel}$ $\gets$ \textbf{MPC-Skeleton}($\mathnormal{Y}$) \hfill \textit{\% using Eq. 3-5}
        \State $\mathcal{L}$ $\gets$ $\mathcal{L}_{generic} (\mathnormal{Y}, \hat{\mathnormal{Y}}) $ + $\mathcal{L}_{skel-weight}(\mathnormal{Y}, \mathnormal{Y}_{MPC-Skel}, \hat{\mathnormal{Y}})$
        \State \Return $FDPS_{\{x,y,z\}}$, $\mathcal{L}$ 
    \end{algorithmic}
\end{algorithm}

\makeatletter
\def\hlinew#1{%
\noalign{\ifnum0=`}\fi\hrule \@height #1 \futurelet
\reserved@a\@xhline}
\makeatother
\begin{table}[t]
\renewcommand\arraystretch{1}
\caption{Quantitative results on \textbf{AortaSeg24 dataset}. `F' denotes FDPS, and `M' denotes MPC-Skel. Dice(\%), clDice(\%), Hd95($mm$), and $\beta_{0}$ error are reported.}\label{tab:results_on_aorta24}
\centering
\scalebox{1.0}{
\begin{tabular}{>{\centering\arraybackslash}p{2.1cm}|>{\centering\arraybackslash}p{0.5cm}|>{\centering\arraybackslash}p{0.5cm}|>{\centering\arraybackslash}p{2.0cm}|>{\centering\arraybackslash}p{2.0cm}|>{\centering\arraybackslash}p{2.0cm}|>{\centering\arraybackslash}p{2.0cm}}
\hlinew{0.8pt}
\multicolumn{3}{c|}{\textbf{Methodology}} & \multicolumn{4}{c}{\textbf{Metrics}} \\ \hlinew{0.8pt}
\textbf{Backbone} & \textbf{F} & \textbf{M} & \textbf{Dice $\uparrow$} & \textbf{clDice $\uparrow$} & \textbf{Hd95 $\downarrow$} & \textbf{$\beta_{0}$ Error $\downarrow$} \\ \hlinew{0.8pt}
\multirow{4}{*}{\makecell{nnUNet \cite{isensee2021nnu} w/ \\ SR \cite{SkeletonRecall}}}    &  \ding{55}  & \ding{55} & 74.78 $\pm$ 4.67 & 97.06 $\pm$ 1.49  & 15.17 $\pm$ 16.45 & 0.50 $\pm$ 0.49 \\ \cline{2-7} 
                                                                                                        &  \ding{51}  & \ding{55} & 76.71 $\pm$ 4.03 & 97.57 $\pm$ 1.23  & 7.55 $\pm$ 3.95 & 0.25 $\pm$ 0.24  \\ \cline{2-7} 
                                                                                                        &  \ding{55}  & \ding{51} & 75.23 $\pm$ 4.82 & 97.27 $\pm$ 1.36  & 10.32 $\pm$ 8.52 & 0.28 $\pm$ 0.19 \\ \cline{2-7} 
                                                                                                        &  \ding{51}  & \ding{51} & \textbf{77.20 $\pm$ 4.30} & \textbf{97.71 $\pm$ 0.89} & \textbf{6.05 $\pm$ 1.76}  & \textbf{0.21 $\pm$ 0.19} \\ \hlinew{0.8pt}
\multirow{4}{*}{\makecell{nnUNet \cite{isensee2021nnu} w/ \\ cbDice \cite{shi2024centerline}}}          &  \ding{55}  & \ding{55} & 74.61 $\pm$ 4.63 & 97.75 $\pm$ 0.82  & 7.22 $\pm$ 2.54 &  0.30 $\pm$ 0.33  \\ \cline{2-7} 
                                                                                                        &  \ding{51}  & \ding{55} & 76.66 $\pm$ 3.87 & 97.80 $\pm$ 1.17  & 6.97 $\pm$ 3.12 & 0.19 $\pm$ 0.14 \\ \cline{2-7} 
                                                                                                        &  \ding{55}  & \ding{51} & 75.26 $\pm$ 4.48 & 97.83 $\pm$ 1.25  & 7.18 $\pm$ 2.32 & 0.23 $\pm$ 0.20 \\ \cline{2-7} 
                                                                                                        &  \ding{51}  & \ding{51} & \textbf{76.80 $\pm$ 3.75} & \textbf{97.83 $\pm$ 0.76} & \textbf{6.41 $\pm$ 1.52} & \textbf{0.14 $\pm$ 0.14} \\ \hlinew{0.8pt}
\multirow{4}{*}{\makecell{nnUNet \cite{isensee2021nnu} w/ \\ CAL \cite{zhang2023towards}}}              &  \ding{55}  & \ding{55} & 75.28 $\pm$ 4.63 & 97.20 $\pm$ 2.27 & 8.87 $\pm$ 7.20 & 0.12 $\pm$ 0.08 \\ \cline{2-7} 
                                                                                                        &  \ding{51}  & \ding{55} & 76.64 $\pm$ 3.90 & \textbf{97.55 $\pm$ 1.49}  & 6.21 $\pm$ 1.29 & 0.09 $\pm$ 0.08 \\ \cline{2-7} 
                                                                                                        &  \ding{55}  & \ding{51} & 75.63 $\pm$ 4.76 & 97.52 $\pm$ 1.53 & 7.06 $\pm$ 2.41 & 0.11 $\pm$ 0.08 \\ \cline{2-7} 
                                                                                                        &  \ding{51}  & \ding{51} & \textbf{77.55 $\pm$ 3.52} & 97.53$\pm$ 1.23  & \textbf{5.67 $\pm$ 1.41} & \textbf{0.07 $\pm$ 0.05} \\ \hlinew{0.8pt}

\end{tabular}}
\end{table}

\subsection{Shape-aware Sampling: Minimum Path-cost Skeleton Weighting}
The SAS introduces the Minimum Path-cost Skeleton Weighting (MPC-Skel) to generate topology-preserved skeletal representation. As illustrated in Fig.\ref{fig:framework}.(b), 
MPC-Skel aims to reduce the clump-like or spurious noises caused by conventional skeletonization techniques \cite{lee1994building, shit2021cldice}. 
Specifically, given a volumetric shape $\mathcal{S}$, a seed point is firstly selected within $\mathcal{S}$, 
and the shortest path is iteratively sampled to sequentially reach the farthest unvisited voxel. The shortest path is determined by a cost function based on the distance transformation: 
\begin{align}
&\mathrm{Dist}(s)  = \min_{\forall \varphi \in \partial \mathcal{S}} d (s,\varphi), \label{eq:dist_func}  \\
&C(s)  = \alpha_{1}(1-\frac{\mathrm{Dist}(s)}{\max_{\forall p \in \mathcal{S}} \mathrm{Dist}(p)})^{\gamma}, \label{eq:cost_func}
\end{align}
where the $\mathrm{Dist}(s)$ measures the shortest distance from an interior voxel $s$ to the surface $\partial \mathcal{S}$ of the shape $\mathcal{S}$. $d(\cdot, \cdot)$ denotes the Euclidean distance. 
The cost function $C(s)$ assigns minimal values to voxels near the center and maximal values to those near the boundary. 
The unvisited voxel with the largest cost function within $\mathcal{S}$ is assigned as the seed point for each path. Starting from the seed point and using the 26-neighbourhood connectivity, 
the shortest path is sampled iteratively, minimizing the $C(s)$ until the farthest unvisited voxel is reached. This minimum path-cost approach helps to mitigate clump-like noise.
After each pass, an adaptive sphere is applied to expand the set of voxels belonging to the path, marking those near the path. 
The adaptive radius $R(s)$ is deterimined by the linear combination of the $\mathrm{Dist}(s)$:
\begin{align}
R(s) = \alpha_{2} \cdot \mathrm{Dist}(s) + \beta. \label{eq:radius_func}
\end{align}
The voxels within the $R(s)$ are labeled visited on the path. The combination of $\alpha_{2}$ and $\beta$ determines the minimum feature size for significant skeletons, which aids in eliminating spurious noise. 
The full skeleton is obtained by tracing all paths until the entire volume is traversed. 
MPC-Skel produces a more accurate skeleton-weighted map and directs the network to focus on critical topological regions, which is beneficial to enhancing tubular topology preservation.


\makeatletter
\def\hlinew#1{%
\noalign{\ifnum0=`}\fi\hrule \@height #1 \futurelet
\reserved@a\@xhline}
\makeatother
\begin{table}[t]
\renewcommand\arraystretch{1}
\caption{Results on \textbf{Airway Anatomical Labeling dataset}. `F' denotes FDPS, and `M' denotes MPC-Skel. Dice(\%), clDice(\%), Hd95($mm$), $\beta_{0}$ error are reported.}\label{tab:results_on_ATMplusplus}
\centering
\scalebox{1.0}{
\begin{tabular}{>{\centering\arraybackslash}p{2.1cm}|>{\centering\arraybackslash}p{0.5cm}|>{\centering\arraybackslash}p{0.5cm}|>{\centering\arraybackslash}p{2.0cm}|>{\centering\arraybackslash}p{2.0cm}|>{\centering\arraybackslash}p{2.0cm}|>{\centering\arraybackslash}p{2.0cm}}
\hlinew{0.8pt}
\multicolumn{3}{c|}{\textbf{Methodology}} & \multicolumn{4}{c}{\textbf{Metrics}} \\ \hlinew{0.8pt}
\textbf{Backbone} & \textbf{F} & \textbf{M} & \textbf{Dice $\uparrow$} & \textbf{clDice $\uparrow$} & \textbf{Hd95 $\downarrow$} & \textbf{$\beta_{0}$ Error $\downarrow$} \\ \hlinew{0.8pt}
\multirow{4}{*}{\makecell{nnUNet \cite{isensee2021nnu} w/ \\ SR \cite{SkeletonRecall}}}    &  \ding{55}  & \ding{55} & 81.05 $\pm$ 5.10 & 87.61 $\pm$ 3.87 & 6.68 $\pm$ 2.24  & 3.91 $\pm$ 1.38 \\ \cline{2-7} 
                                                                                                        &  \ding{51}  & \ding{55} & 83.30 $\pm$ 4.86 & 89.86 $\pm$ 3.12 & \textbf{4.91 $\pm$ 2.25} & 3.54 $\pm$ 1.11 \\ \cline{2-7} 
                                                                                                        &  \ding{55}  & \ding{51} & 82.08 $\pm$ 5.95 & 89.00 $\pm$ 3.28 & 6.21 $\pm$ 2.49  & 3.78 $\pm$ 1.26  \\ \cline{2-7} 
                                                                                                        &  \ding{51}  & \ding{51} & \textbf{84.08 $\pm$ 3.85} & \textbf{89.92 $\pm$ 1.73} & 4.93 $\pm$ 1.82  & \textbf{3.38 $\pm$ 1.02} \\ \hlinew{0.8pt}
\multirow{4}{*}{\makecell{nnUNet \cite{isensee2021nnu} w/ \\ cbDice \cite{shi2024centerline}}}          &  \ding{55}  & \ding{55} & 83.19 $\pm$ 5.19 & 89.34 $\pm$ 3.03 & 5.60 $\pm$ 2.22  & 2.47 $\pm$ 0.91  \\ \cline{2-7}
                                                                                                        &  \ding{51}  & \ding{55} & 83.80 $\pm$ 5.21 & 89.79 $\pm$ 2.71 & \textbf{5.35 $\pm$ 2.40} & 2.38 $\pm$ 0.85 \\ \cline{2-7}
                                                                                                        &  \ding{55}  & \ding{51} & 83.62 $\pm$ 6.00 & \textbf{89.94 $\pm$ 2.63} & 5.75 $\pm$ 2.33 & \textbf{2.28 $\pm$ 0.78}   \\ \cline{2-7}
                                                                                                        &  \ding{51}  & \ding{51} & \textbf{84.10 $\pm$ 4.85} & 89.92 $\pm$ 2.54 & 5.56 $\pm$ 2.49 & 2.32 $\pm$ 0.72  \\ \hlinew{0.8pt}
\multirow{4}{*}{\makecell{nnUNet \cite{isensee2021nnu} w/ \\ CAL \cite{zhang2023towards}}}              &  \ding{55}  & \ding{55} & 83.63 $\pm$ 5.57 & 89.69 $\pm$ 3.15 & 5.82 $\pm$ 2.36  & 3.43 $\pm$ 1.38  \\ \cline{2-7}
                                                                                                        &  \ding{51}  & \ding{55} & 84.13 $\pm$ 5.23 & \textbf{89.94 $\pm$ 2.76} & 5.68 $\pm$ 2.35  & \textbf{2.98 $\pm$ 0.84}  \\ \cline{2-7} 
                                                                                                        &  \ding{55}  & \ding{51} & 84.00 $\pm$ 5.44 & 89.58 $\pm$ 2.97 & 5.79 $\pm$ 2.08  & 3.21 $\pm$ 0.87  \\ \cline{2-7} 
                                                                                                        &  \ding{51}  & \ding{51} & \textbf{84.38 $\pm$ 4.68} & 89.83 $\pm$ 2.61 & \textbf{5.11 $\pm$ 2.14}  & 3.18 $\pm$ 1.01  \\ \hlinew{0.8pt}

\end{tabular}}
\end{table}

\begin{figure*}[t]
\centering
\includegraphics[width=0.80\linewidth]{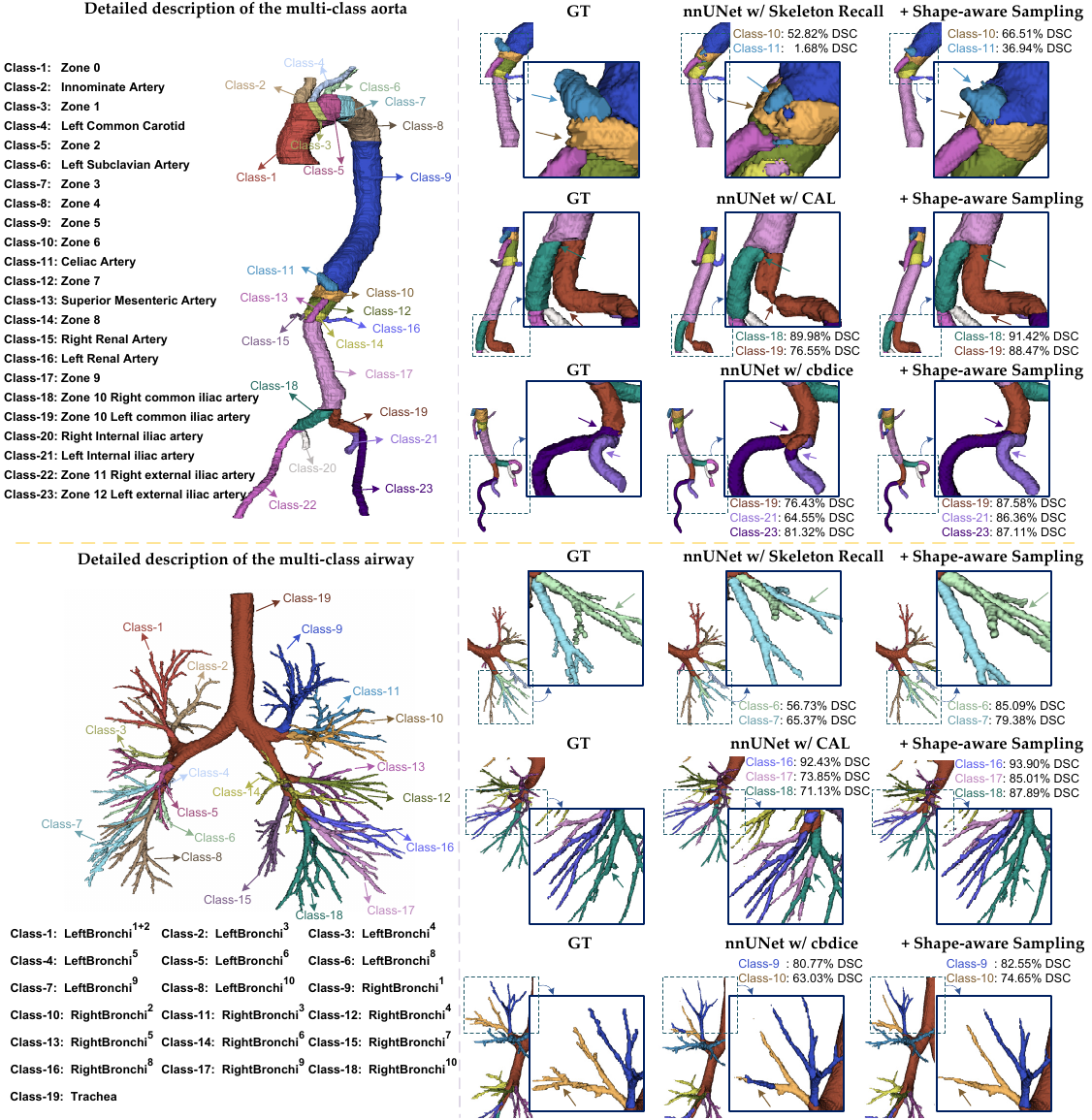}
\caption{Visualization of the results on the modeling of multi-class aorta and airway.}
\label{fig:Visual_Result_1}
\end{figure*}

\section{Experiments and Results}
\noindent \textbf{Dataset}:
This study incorporates two multi-class tubular datasets. 1) \textbf{AortaSeg24 dataset} \cite{imran2025multi}, which comprises 50 CTA volumes featuring 23 clinically significant aortic branches and zones. 
The image volumes were uniformly resampled to an isotropic resolution of $1\,\text{mm} \times 1\,\text{mm} \times 1\,\text{mm}$, with an average image dimension of $441 \times 441 \times 686$ in the $x \times y \times z$ axes.  
2) \textbf{Airway Anatomical Labeling dataset} \cite{yu2022tnn}, which includes 104 chest CT scans. These scans have a slice thickness of $0.67\,\text{mm}$, with the $x$ and $y$ axes resolutions 
ranging from $0.78\,\text{mm}$ to $0.82\,\text{mm}$. This dataset employs segmental anatomical labeling of the airway, including eight anatomical bronchi in the left lung, ten branches in the right lung, and one trachea. 
A qualitative depiction of the multi-class aorta and airway structures is illustrated in Fig.~\ref{fig:Visual_Result_1}. Both datasets were randomly split into training, validation, and test sets in a 6:1:3 ratio.\\
\noindent \textbf{Implementation Details}: 
All experiments were performed on the nnUNet framework \cite{isensee2021nnu}. Mirror augmentation was disabled to avoid any positional confusion. 
The axis-specific fractal dimensions were $\{0.58, 0.58, 0.71\}$ for the multi-class aorta, with the initial patchsize configuration set to $\{112, 112, 176\}$ along the $x \times y \times z$ axes. 
According to Eq.\ref{eq:FDPS}, the resulting FDPS configuration for AortaSeg24 is $\{176, 176, 112\}$. To ensure a fair comparison, the original patchsize was increased to ${288, 112, 112}$. 
For the multi-class airway, the axis-specific fractal dimensions were $\{0.44, 0.40, 0.53\}$, with the patchsize adjusted from $\{128, 96, 192\}$ to $\{128, 192, 96\}$ following the FDPS approach. 
Three effective backbones, which incorporate skeleton-weighted loss funcition \cite{zhang2023towards,SkeletonRecall,shi2024centerline}, were implemented in this work. 
All backbones follow the loss formulation: $\mathcal{L} = \mathcal{L}_{generic} + \mathcal{L}_{skel-weight}$.
The $\mathcal{L}_{generic}$ denotes the combination of Cross-Entropy and Soft Dice loss. In the $ \mathcal{L}_{skel-weight}$ term, 
the skeletonization techniques \cite{lee1994building,shit2021cldice} were replaced with the MPC-Skel. 
Inspired by \cite{sato2000teasar, bitter2001penalized}, The calculation of MPC-Skel is based on accumulated path distance, which ensures computational efficiency.
In all experiments, $\alpha_{1} = 1e5$ and $\gamma = 4$ were used. For the multi-class aorta, 
$\alpha_{2} = 1.8 \times \mathrm{spacing_{max}}$ and $\beta = 4 \times \mathrm{spacing_{max}}$ were set, while for the multi-class airway, 
$\alpha_{2} = 2.4 \times \mathrm{spacing_{max}}$ and $\beta = 2 \times \mathrm{spacing_{max}}$ were applied. Here, $\mathrm{spacing_{max}}$ denotes the maximum resolution spacing among the $\{x, y, z\}$ axes.

\begin{figure*}[t]
\centering
\includegraphics[width=0.85\linewidth]{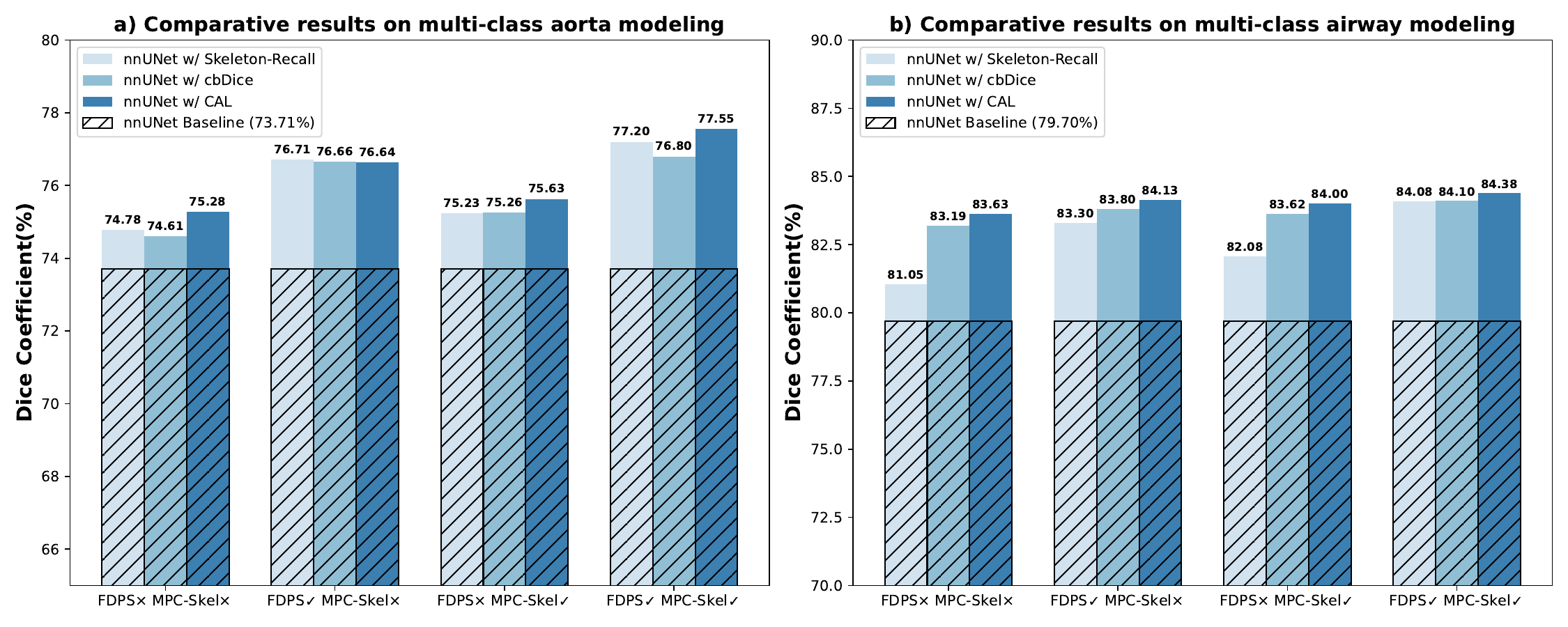}
\caption{Comparative results with baselines for multi-class aorta and airway modeling.}
\label{fig:Visual_Result_2}
\end{figure*}

\noindent \textbf{Quantitative Results Analysis}: Four metrics were used to assess both volumetric overlap and topological integrity: 1) Dice score coefficient (Dice, \%). 
2) 95\% Hausdorff distance (Hd95, $mm$). 3) clDice \cite{shit2021cldice}  (\%). 4) $\beta_{0}$ Error. 
As illustrated in Fig.\ref{fig:Visual_Result_2}, the backbones incorporating skeleton-weighted loss functions showed improvements over the baseline nnUNet framework.
Specifically, the baseline achieved average Dice of 73.71\% and 79.70\% on the AortaSeg24 and Airway Anatomical Labeling dataset, respectively. 
With the integration of Skeleton-Recall (SR) \cite{SkeletonRecall}, cbDice \cite{shi2024centerline}, and CAL \cite{zhang2023towards}, 
the performance increased to 74.78\%/74.61\%/75.28\% on the multi-class aorta, and 81.05\%/83.19\%/83.63\% on multi-class airway. 
When the proposed SAS framework was applied to these three backbones, it consistently improve the performance across both datasets.
As detailed in Tab.\ref{tab:results_on_aorta24} and Tab.\ref{tab:results_on_ATMplusplus}, 
SAS improves the average Dice score by over 2\% on the AortaSeg24 dataset on all three backbones. 
It achieves approximately 84\% Dice on the multi-class airway, representing an increase of around 4.5\% compared to the baseline.
Furthermore, SAS demonstrates its capability in enhancing topological integrity. For example, with the nnUNet incorporating Skeleton-Recall \cite{SkeletonRecall}, 
the FDPS reduces the Hd95 error from 15.17 $mm$ to 7.55 $mm$, and decreases the $\beta_{0}$ error from 0.5 to 0.25 on AortaSeg24. When combined with MPC-Skel, the Hd95 and $\beta_{0}$ errors 
are further optimized to 6.05 $mm$ and 0.21, respectively. 
Similar trends were observed across other experimental configurations. Notably, while nnUNet incorporating CAL \cite{zhang2023towards} achieved a comparative 5.82 $mm$ Hd95 and 
3.43 $\beta_{0}$ error without SAS, the introduction of SAS further enhanced performance, reducing the Hd95 to 5.11 $mm$ and the $\beta_{0}$ error to 3.18.

\noindent \textbf{Qualitative Results Analysis}: 
Fig.\ref{fig:Visual_Result_1} visualizes the improvement achieved by SAS. 
As shown in the first row of the comparison within the AortaSeg24 dataset, Zone 6, the Celiac Artery, Zone 7, and the Superior Mesenteric Artery are in close proximity to each other. 
This area exhibits high shape complexity along the z-axis. The method with SAS models these structures more accurately, owing to its refined patchsize reallocation and the enhanced skeleton-weighted map, 
which better capture the intricacies of the complex structures. The arrows highlight how SAS improves the topological integrity of the Zone 6 and Celiac Artery structures. 
For the multi-class airway, SAS consistently enhances topological integrity across different backbones. The highlighted regions indicate that SAS better preserves the topological consistency of different classes of bronchi.

\section{Conclusion}
This paper introduces the Shape-aware Sampling (SAS) to improve multi-class tubular structures modeling. SAS consists of two efficient components: the Fractal Dimension-based Patchsize (FDPS) and Minimum Path-Cost Skeletonization (MPC-Skel). 
FDPS optimizes patchsize allocation for online sampling based on axis-specific fractal dimensions, enhancing the capture of complex shapes. 
MPC-Skel refines skeleton-weighted maps by generating topology-preserved representations. 
Experimental results show that SAS integration significantly improves both volumetric overlap and topological integrity.

\bibliography{paper.bib}

\begin{thebibliography}{10}
\providecommand{\url}[1]{\texttt{#1}}
\providecommand{\urlprefix}{URL }
\providecommand{\doi}[1]{https://doi.org/#1}

\bibitem{attali2009stability}
Attali, D., Boissonnat, J.D., Edelsbrunner, H.: Stability and computation of
  medial axes-a state-of-the-art report. Mathematical foundations of scientific
  visualization, computer graphics, and massive data exploration pp. 109--125
  (2009)

\bibitem{bitter2001penalized}
Bitter, I., Kaufman, A.E., Sato, M.: Penalized-distance volumetric skeleton
  algorithm. IEEE Transactions on Visualization and computer Graphics
  \textbf{7}(3),  195--206 (2001)

\bibitem{carrel2023acute}
Carrel, T., Sundt, T.M., von Kodolitsch, Y., Czerny, M.: Acute aortic
  dissection. The Lancet  \textbf{401}(10378),  773--788 (2023)

\bibitem{cciccek20163d}
{\c{C}}i{\c{c}}ek, {\"O}., Abdulkadir, A., Lienkamp, S.S., Brox, T.,
  Ronneberger, O.: 3d u-net: learning dense volumetric segmentation from sparse
  annotation. In: Medical Image Computing and Computer-Assisted
  Intervention--MICCAI 2016: 19th International Conference, Athens, Greece,
  October 17-21, 2016, Proceedings, Part II 19. pp. 424--432. Springer (2016)

\bibitem{cornea2024curve}
Cornea, N.D., Silver, D., Min, P.: Curve-skeleton properties, applications, and
  algorithms. IEEE Transactions on visualization and computer graphics
  \textbf{13}(3),  530--548 (2024)

\bibitem{falconer2013fractal}
Falconer, K.: Fractal geometry: mathematical foundations and applications. John
  Wiley \& Sons (2013)

\bibitem{imran2024cis}
Imran, M., Krebs, J.R., Gopu, V.R.R., Fazzone, B., Sivaraman, V.B., Kumar, A.,
  Viscardi, C., Heithaus, R.E., Shickel, B., Zhou, Y., et~al.: Cis-unet:
  Multi-class segmentation of the aorta in computed tomography angiography via
  context-aware shifted window self-attention. Computerized Medical Imaging and
  Graphics  \textbf{118},  102470 (2024)

\bibitem{imran2025multi}
Imran, M., Krebs, J.R., Sivaraman, V.B., Zhang, T., Kumar, A., Ueland, W.R.,
  Fassler, M.J., Huang, J., Sun, X., Wang, L., et~al.: Multi-class segmentation
  of aortic branches and zones in computed tomography angiography: The
  aortaseg24 challenge. arXiv preprint arXiv:2502.05330  (2025)

\bibitem{isensee2021nnu}
Isensee, F., Jaeger, P.F., Kohl, S.A., Petersen, J., Maier-Hein, K.H.: nnu-net:
  a self-configuring method for deep learning-based biomedical image
  segmentation. Nature methods  \textbf{18}(2),  203--211 (2021)

\bibitem{isensee2024scaling}
Isensee, F., Kirchhoff, Y., Kraemer, L., Rokuss, M., Ulrich, C., Maier-Hein,
  K.H.: Scaling nnu-net for cbct segmentation. arXiv preprint arXiv:2411.17213
  (2024)

\bibitem{isensee2024nnu}
Isensee, F., Wald, T., Ulrich, C., Baumgartner, M., Roy, S., Maier-Hein, K.,
  Jaeger, P.F.: nnu-net revisited: A call for rigorous validation in 3d medical
  image segmentation. In: International Conference on Medical Image Computing
  and Computer-Assisted Intervention. pp. 488--498. Springer (2024)

\bibitem{SkeletonRecall}
Kirchhoff, Y., Rokuss, M.R., Roy, S., Kovacs, B., Ulrich, C., Wald, T., Zenk,
  M., Vollmuth, P., Kleesiek, J., Isensee, F., et~al.: Skeleton recall loss for
  connectivity conserving and resource efficient segmentation of thin tubular
  structures. In: European Conference on Computer Vision. pp. 218--234.
  Springer (2024)

\bibitem{konatar2020box}
Konatar, I., Popovic, T., Popovic, N.: Box-counting method in python for
  fractal analysis of biomedical images. In: 2020 24th International Conference
  on Information Technology (IT). pp.~1--4. IEEE (2020)

\bibitem{lee1994building}
Lee, T.C., Kashyap, R.L., Chu, C.N.: Building skeleton models via 3-d medial
  surface axis thinning algorithms. CVGIP: Graphical Models and Image
  Processing  \textbf{56}(6),  462--478 (1994)

\bibitem{li2009improved}
Li, J., Du, Q., Sun, C.: An improved box-counting method for image fractal
  dimension estimation. Pattern recognition  \textbf{42}(11),  2460--2469
  (2009)

\bibitem{liebovitch1989fast}
Liebovitch, L.S., Toth, T.: A fast algorithm to determine fractal dimensions by
  box counting. physics Letters A  \textbf{141}(8-9),  386--390 (1989)

\bibitem{pentland1984fractal}
Pentland, A.P.: Fractal-based description of natural scenes. IEEE transactions
  on pattern analysis and machine intelligence (6),  661--674 (1984)

\bibitem{rolf2024mechanisms}
Rolf-Pissarczyk, M., Schussnig, R., Fries, T.P., Fleischmann, D., Elefteriades,
  J.A., Humphrey, J.D., Holzapfel, G.A.: Mechanisms of aortic dissection: from
  pathological changes to experimental and in silico models. Progress in
  Materials Science p. 101363 (2024)

\bibitem{sato2000teasar}
Sato, M., Bitter, I., Bender, M.A., Kaufman, A.E., Nakajima, M.: Teasar:
  tree-structure extraction algorithm for accurate and robust skeletons. In:
  Proceedings the Eighth Pacific Conference on Computer Graphics and
  Applications. pp. 281--449. IEEE (2000)

\bibitem{shi2024centerline}
Shi, P., Hu, J., Yang, Y., Gao, Z., Liu, W., Ma, T.: Centerline boundary dice
  loss for vascular segmentation. In: International Conference on Medical Image
  Computing and Computer-Assisted Intervention. pp. 46--56. Springer (2024)

\bibitem{shit2021cldice}
Shit, S., Paetzold, J.C., Sekuboyina, A., Ezhov, I., Unger, A., Zhylka, A.,
  Pluim, J.P., Bauer, U., Menze, B.H.: cldice-a novel topology-preserving loss
  function for tubular structure segmentation. In: Proceedings of the IEEE/CVF
  Conference on Computer Vision and Pattern Recognition. pp. 16560--16569
  (2021)

\bibitem{yu2022tnn}
Yu, W., Zheng, H., Gu, Y., Xie, F., Yang, J., Sun, J., Yang, G.Z.: Tnn: Tree
  neural network for airway anatomical labeling. IEEE Transactions on Medical
  Imaging  \textbf{42}(1),  103--118 (2022)

\bibitem{zhang2023soft}
Zhang, J., Fang, Q., Xiang, P., Xiong, R., Wang, Y., Lu, H.: Soft hybrid
  actuated hierarchical bronchoscope robot for deep lung examination. IEEE
  Robotics and Automation Letters  \textbf{9}(1),  811--818 (2023)

\bibitem{zhang2023towards}
Zhang, M., Gu, Y.: Towards connectivity-aware pulmonary airway segmentation.
  IEEE Journal of Biomedical and Health Informatics  (2023)

\end{thebibliography}
\end{document}